# Dammann metasurface route to overcoming the uniformity defects in two-dimensional beam multipliers

Rinat Gutin[1†], Raghvendra P. Chaudhary[1†], Imon Kalyan[1], Avraham Reiner[1], and Nir Shitrit[1*]

[1]*School of Electrical and Computer Engineering, Ben-Gurion University of the Negev, Be'er Sheva 8410501, Israel*

(Dated: March 25, 2026)

[†]These authors contributed equally to this work.

[*]Corresponding author. E-mail: nshitrit@bgu.ac.il

## Abstract

Dammann gratings—beam-shaping optical elements acting as beam multipliers with equal-power beams—are a key element in three-dimensional imaging based on structured light and beam combiners for high-power laser applications. However, two-dimensional Dammann grating structures suffer from a significant reduction of the uniformity among the diffraction orders. Here, we report Dammann metasurfaces based on the geometric phase as the structure realization for the target phase profile, which outperform the capabilities of Dammann gratings by overcoming the uniformity defects in their two-dimensional diffraction patterns. We showed that two-dimensional Dammann metasurfaces exhibit high uniformity and diffraction efficiency, in contrast to Dammann gratings, by overcoming the uniformity defects via a robust and highly precise phase imprint. Moreover, Dammann metasurfaces outperform their grating counterparts by exhibiting a polarization-independent response and a broadband operation. This study reveals that by providing physics-driven solutions, metasurfaces can outperform the capabilities of their bulk optics counterparts while facilitating virtually flat, ultrathin, and lightweight optics.

## 1. Introduction

Photonic metasurfaces are two-dimensional (2D) ultrathin arrays of engineered subwavelength-spaced nanoscatterers that mold optical wavefronts at will by imparting phase changes along an interface [1–3]. By providing unprecedented and simultaneous control over the fundamental properties of light—phase, amplitude, and polarization—metasurfaces aim to revolutionize optical designs by realizing virtually flat, ultrathin, and lightweight optics that replaces bulky optical elements [1–34]. The light-scattering properties of metasurfaces can be manipulated through choices of the nanostructures' material, size, geometry, orientation, and environment; therefore, metasurfaces open a new paradigm for flat photonics based on structured interfaces by facilitating one-to-one correspondence between information pixels and nanostructures [1–34]. Metasurfaces consisting of nano-engineered structures have enabled a custom-designed electromagnetic response with unprecedented functionalities in the linear [1–10,12,14,15,18–20,23,29,30,32–34], nonlinear [11,13,16,17,21,24,28], and quantum optical regimes [22,25–27,31]. Owing to its design flexibility and versatility along with a straightforward fabrication, generally requiring single-step lithography, the metasurface platform has become the gold standard for controlling the spatial phase profile of wavefronts with subwavelength sampling [34].

Strikingly, as beam-shaping optical elements, metasurfaces can outperform the capabilities of their bulk refractive and diffractive counterparts. One such example is Dammann gratings (DGs). DGs are diffractive beam-shaping optical elements acting as beam multipliers, i.e., phase-only optical elements that divide an input beam into multiple output beams with equal power [35]. In the context of structured light [36], DGs are a key element as they project a far-field pattern of a uniform array of dots with a high efficiency [37], thus leveraging three-dimensional imaging, facial recognition, etc. [32,38,39]. Laser-wise, DGs play a crucial role as beam combiners (i.e., a beam multiplier in a reversed propagation) which combine multiple lasers in an array configuration [40] to obtain the ever-growing demand for a high-power laser beam. Owing to their ease of fabrication, binary phase DGs are the most common beam-multiplier structures [35,41]. Binary phase DGs are manifested by a complex groove shape, i.e., multiple carefully designed transition points within the period of the grating [35,41,42]. Binary phase $(0,\pi)$ DGs are obtained by etching the grating material to a thickness corresponding to a $\pi$ phase shift. While these DGs exhibit a wavelength-scale thickness, their smallest features are subwavelength [35,42] [see Fig. 1(a) for a typical phase profile of a one-dimensional (1D) DG, which depicts the DG structure]. By combining two 1D DGs in the horizontal and vertical directions, such that their 1D phase maps are summed, the concept of DGs extends to 2D binary gratings; this procedure results in 2D DGs with barcode-like patterns [see Fig. 1(c)]. Application-wise, 2D DGs are the enabling structures. However, while 1D binary phase $(0,\pi)$ DG structures exhibit high uniformity among the diffraction orders, 2D DG structures—constructed from their 1D counterparts—suffer from a significant reduction of the uniformity, as observed in many experiments [38,39,43–46]; moreover, the mechanism for the uniformity defects in 2D DG diffraction patterns [Fig. 1(d)] has remained elusive. To overcome the uniformity challenge in 2D DGs, alternative approaches of structure realizations for a target phase profile should be studied, where metasurfaces are a promising solution for imprinting an abrupt (i.e., rapidly changing) spatial phase in an ultrathin architecture.

Here, we report Dammann metasurfaces (DMs) based on the Pancharatnam–Berry geometric phase that outperform the capabilities of conventional DGs by overcoming the uniformity defects in 2D DG diffraction patterns. We identified the mechanism lowering the uniformity in 2D DG structures as a defective spatial phase imprint differing from the target phase near the transition points with subwavelength smallest features. We have imprinted the target phase via metasurfaces based on the geometric phase, where the phase pickup is enabled by tiling a surface with anisotropic nanogratings arranged according to an on-demand space-variant orientation angle profile. We theoretically and numerically studied 1D DGs and DMs generating 5, 7, and 9 equal-power diffraction orders and their corresponding 2D structures generating 5-by-5, 7-by-7, and 9-by-9 diffraction orders. By calculating the far-field patterns of DGs and DMs and quantifying the uniformity among the diffraction orders, we showed that 2D DMs exhibit high uniformity and diffraction efficiency, in contrast to DGs, by overcoming the uniformity defects via a robust and highly precise phase imprint. Moreover, the DMs outperform their DG counterparts by exhibiting a polarization-independent response and a broadband operation. By providing physics-driven solution for imprinting rapidly changing phase via structured interfaces, the concept of DMs empowers the ubiquitous binary phase optical elements, thus redefining optical design paradigms; moreover, DMs reveal that metasurfaces can outperform the capabilities of their bulk optics counterparts while facilitating virtually flat, ultrathin, and lightweight optics for state-of-the-art wavefront molding at will.

## 2. Methods: Far-field calculation and uniformity analysis of Dammann gratings

The functionality of DGs as beam multipliers with equal-power diffraction orders can be simply illustrated by Fourier analysis. For 1D DGs, the periodic transmission function $t(x) = e^{-i\phi(x)}$, where $\phi(x)$ is the binary phase (0,$\pi$) profile set by the transition points within the period of the grating $P$ [Fig. 1(a)], is analyzed in terms of Fourier series. We express the Fourier decomposition (i.e., the angular spectrum) of a DG by calculating the Fourier series coefficients of the transmission function $c_n = \frac{1}{P}\int_0^P t(x)e^{-i2\pi nx/P}dx$, where $n$ is the harmonic order—i.e., the diffraction order. The intensities $|c_n|^2$ reveal the equal-power diffraction orders along with their high efficiency [Fig. 1(b)]. To rigorously quantify the uniformity among the diffraction orders of DGs, we extend the Fourier toy model by theoretically calculating the DG far-field diffraction patterns based on the corresponding binary phase (0,$\pi$) profile. By considering a monochromatic, unit-amplitude, normally incident plane-wave illumination, we expressed the electric field at the grating $x$–$y$ plane as $E_0(x,y,z = 0) = e^{i\phi(x,y)}$, where $\phi(x,y) = \phi_x + \phi_y$; $\phi_x$ and $\phi_y$ are the 1D phase maps in the $x$ and $y$ directions, respectively [see Fig. 1(c)]. By employing the Huygens–Fresnel diffraction integral [22,47], the scalar field at the observation $x'$–$y'$ plane is calculated as

$$E_1(x',y',z) = i\frac{k}{2\pi}\iint \cos\theta\left(1+\frac{i}{kr}\right)\frac{e^{ikr}}{r}E_0(x,y,z=0)dxdy \tag{1}$$

where $r = \sqrt{(x-x')^2+(y-y')^2+z^2}$, $z$ is the distance between the grating plane and the observation plane, $\cos\theta = z/r$ is the inclination factor, and $k$ is the wave number. We calculated the DG far-field diffraction patterns while satisfying the Fraunhofer diffraction condition of $z \gg \pi m^2P^2/2\lambda$, where $\lambda$ is the wavelength, $P$ is the period of the grating structure, and $m$ is the number of the grating replicas (the size of the grating structure is $mP$). We consider a DG with a period $P$ of 20 μm and a total size of 200 μm ($m = 10$), at the wavelength of 630 nm; the transition points within the period of the grating were determined based on an optimization maximizing the uniformity and the efficiency of the desired diffraction orders, reported elsewhere [42]. Figure 2(a) shows a typical 2D DG phase map, which specifically generates 5-by-5 equal-power diffraction orders, whereas Fig. 2(b) reveals the corresponding far-field intensity pattern. To quantify the uniformity among the diffraction orders, we calculated for each diffraction order the intensity integrated over an area (i.e., power) of a diffraction-limited spot with a size of $2\lambda z/mP$. We defined the uniformity of the equal-power diffraction orders as

$$U = 1 - \frac{\max(I_{i,j}) - \min(I_{i,j})}{\max(I_{i,j}) + \min(I_{i,j})} \tag{2}$$

where $I_{i,j}$ is the integrated intensity (power) of the $(i,j)$ diffraction order. Moreover, the diffraction efficiency is defined as the sum of the relative power of the desired diffraction orders. The calculated uniformities of the DGs based on their phase profiles exceed 99% [see Fig. 2(b)], revealing the diffraction orders are extremely uniform. Note that this far-field calculation is purely based on the binary phase $(0,\pi)$ profile—without considering a structure realization for the target phase. Nevertheless, the behavior of DGs based on physical structure realizations may change significantly. Figure 2(c) shows the typical structure realization (single period) of a 2D binary phase $(0,\pi)$ DG corresponding to 5-by-5 equal-power diffraction orders, which is obtained by etching the grating material (fused silica) to a thickness corresponding to a $\pi$ phase shift. We numerically simulated the near-field phase profile of this DG structure [Fig. 2(e)] and the corresponding far-field diffraction pattern [Fig. 2(g)]. It is evident that the imprinted near-field phase [Fig. 2(e)] deviates from the target phase [Fig. 2(a)], giving rise to a defective spatial phase imprint induced by the structure realization. While the binary phase $(0,\pi)$ DG structures exhibit a wavelength-scale thickness, the phase defects are dominant near the transition points with subwavelength smallest features; moreover, the imprinted phase is nonuniform, thus deviating from the target of binary phase $(0,\pi)$ [see Fig. 2(i)]. This defective spatial phase imprint induces significant reduction in the uniformity among the diffraction orders in the far-field pattern, where for 5-by-5 diffraction orders the uniformity reduces to 66% [Fig. 2(g)]. Such uniformity defects in the far-field diffraction pattern of DG structures were also observed in many experiments [38,39,43–46]; however, the mechanism for these uniformity defects has not yet been identified.

## 3. Results

### 3.1. The role of the structure realization: Overcoming the uniformity defects in two-dimensional Dammann gratings via metasurfaces

To overcome the uniformity challenge in 2D DGs via a robust and highly precise phase imprint, we consider DMs based on the Pancharatnam–Berry geometric phase [48–50] as the structure realization for the target phase profile. The pickup of the Pancharatnam–Berry geometric phase in metasurfaces arises from the space-variant manipulation of the polarization state of light, associated with a closed loop traverse upon the Poincaré sphere [1,10,48,50], that is enabled by tiling a surface with anisotropic nanoscatterers arranged according to an on-demand space-variant orientation angle profile $\theta(x,y)$ [1,4–6,10,12,18–20,25,26]. Particularly, a metasurface based on the geometric phase converts an incident circularly polarized light into a beam of opposite circular polarization, imprinted with a geometric phase $\phi_g(x,y) = -2\sigma_{\pm}\theta(x,y)$, where $\sigma_{\pm} = \pm 1$ is the polarization helicity of the incident light corresponding to right and left circularly polarized light, respectively [6,9,19,20,25,26]. In contrast to a propagating (dynamic) phase emerging from optical path differences, the geometric phase is manifested by a geometric origin, where this phase imprint concept has been providing a versatile platform for helicity-dependent metasurface elements with desired functionalities [1,4–6,9,10,12,18–20,25,26]. Figure 2(d) shows the structure realization (single period) of a 2D binary phase $(0,\pi)$ DM corresponding to 5-by-5 equal-power diffraction orders, obtained by tiling polycrystalline silicon (poly-Si) nanobeams according to a binary orientation angle profile $\theta(x,y) = \phi(x,y)/2$, where $\phi(x,y)$ is the target phase map [Fig. 2(a)]. The numerically simulated near-field phase of the DM [Fig. 2(f)] reveals a highly precise and uniform phase imprint compared to DGs [Fig. 2(e)], which yields high uniformity of 96% among the diffraction orders in the DM far-field pattern [Fig. 2(h)]. To validate that DMs outperform the performance of DGs by overcoming the uniformity defects in the 2D diffraction patterns, we quantified the mismatch between the target and imprinted phases via the root-mean-square error

$$\mathrm{RMSE} = \sqrt{\frac{\sum_{i=1}^{n}\left(\phi_i - \hat{\phi}_i\right)^2}{n}} \tag{3}$$

where $\phi_i$ and $\hat{\phi}_i$ are the local imprinted and target phases, respectively, and $n$ is the number of data points. The calculated RMSEs for DMs [Fig. 2(j)] exhibit lower values compared to DGs [Fig. 2(i)] revealing the robust and highly precise phase imprint of metasurfaces based on the geometric phase as the structure realization for a target phase profile; this trend in the RMSEs reflects on the higher uniformities observed in DM far-field patterns [Fig. 2(h)], in comparison with DGs [Fig. 2(g)]. These observations reveal the one-to-one correspondence between the defective spatial phase imprint differing from the target phase and the uniformity defects in the far-field diffraction patterns of beam multipliers.

To further support that metasurfaces based on the geometric phase empower binary phase optical elements, we studied DGs and DMs corresponding to various equal-power diffraction orders. Figures 3(a)–3(c) show the 1D phase maps (unit cells) corresponding to 5, 7, and 9 equal-power diffraction orders, respectively, whereas Figs. 3(d)–3(f) present the corresponding calculated far-field patterns. Similarly, Figs. 3(g) and 3(h) show the 2D phase maps (unit cells) corresponding to 7-by-7 and 9-by-9 desired diffraction orders, respectively, where Figs. 3(i) and 3(j) exhibit the corresponding calculated far-field patterns. The carefully designed transition points within the period of the grating were reported elsewhere [42]. These 1D and 2D far-field calculations—based on the perfect binary phase $(0,\pi)$ profile which disregards a structure realization for the target phase—exhibit near-unity uniformities among the equal-power diffraction orders for all DG designs [Fig. 3(k)]. However, the actual behavior of DGs based on physical structure realizations considerably differs from the theoretical expectations (Fig. 3). We realized binary phase $(0,\pi)$ DGs based on the target phase maps [Figs. 3(a)–3(c), 3(g) and 3(h)] by etching the grating material (fused silica) to a thickness corresponding to a $\pi$ phase shift. Figures 4(a)–4(c) depict the 1D DGs (unit cells) corresponding to 5, 7, and 9 desired diffraction orders, respectively, whereas Figs. 4(d)–4(f) show the corresponding simulated far-field patterns for a normally incident horizontal linear polarization, at the wavelength of 630 nm; the size of the grating structures is 200 μm corresponding to 10 replicas of the unit cell, and the far-field patterns correspond to the polarization excitation. Similarly, Figs. 4(g) and 4(h) show the 2D DGs (unit cells) corresponding to 7-by-7 and 9-by-9 desired diffraction orders, respectively, where Figs. 4(i) and 4(j) exhibit the corresponding simulated far-field patterns. The far-field patterns of 1D DGs show high uniformities exceeding 90% [Figs. 4(d)–4(f)]; however, the far-field diffraction patterns of 2D DGs feature uniformity defects, giving rise to extremely low uniformities among the desired diffraction orders—as low as 25% [Fig. 4(j)]. Although 2D DGs are constructed from their 1D counterparts—by summing the 1D phase maps in the $x$ and $y$ directions which generates barcode-like patterns—the uniformity defects are significant in the 2D diffraction patterns compared to the 1D patterns [Fig. 4(k)]; this observation renders that the defective spatial phase imprint is amplified in 2D DGs, enlarging the difference from the target phase. The one-to-one correspondence between the defective spatial phase imprint and the uniformity defects in the far-field diffraction patterns outlines overcoming the uniformity challenge in 2D DGs via a highly precise phase imprint, where we consider metasurfaces as the structure realization for the target phase.

### 3.2. Metasurface design

We consider a metasurface based on the Pancharatnam–Berry geometric phase as a tiling of anisotropic birefringent wave plates (meta-atoms) locally orientated according to the space-variant angle profile $\theta(x,y)$ of their fast axes. For an incident plane wave $|E_{in}\rangle$ with an arbitrary polarization state, the output field $|E_{out}\rangle$ of such a metasurface is composed of three polarization orders

$$|E_{out}\rangle = \sqrt{\eta_E}|E_{in}\rangle + \sqrt{\eta_R}e^{i2\theta(x,y)}|R\rangle + \sqrt{\eta_L}e^{-i2\theta(x,y)}|L\rangle \tag{4}$$

where $|R\rangle$ and $|L\rangle$ are unit Jones vectors of right and left circularly polarized light, respectively [4,5,10]. The quantities $\eta_E = \left|\frac{1}{2}\left(t_x + t_y e^{i\Phi}\right)\right|^2$, $\eta_R = \left|\frac{1}{2}\left(t_x - t_y e^{i\Phi}\right)\langle E_{in}|L\rangle\right|^2$, and $\eta_L = \left|\frac{1}{2}\left(t_x - t_y e^{i\Phi}\right)\langle E_{in}|R\rangle\right|^2$ provide the magnitude of the coupling efficiencies of the $|E_{in}\rangle$, $|R\rangle$, and $|L\rangle$ polarization orders, respectively; we adopt the Dirac bra-ket notation, where $\langle\alpha|\beta\rangle$ denotes an inner product. Moreover, $t_x$ and $t_y$ are the magnitude of the amplitude transmission coefficients for light polarized perpendicular and parallel to the fast optical axis, and $\Phi$ is the birefringent phase retardation between these linear polarizations. Figure 5(a) represents the behavior of a metasurface based on the geometric phase (Eq. [4]); while the $|E_{in}\rangle$ polarization order maintains the polarization and phase of the incident beam, the $|R\rangle$ and $|L\rangle$ polarization orders are accompanied by geometric phase modifications of $\pm 2\theta(x,y)$, respectively [Fig. 5(b)]. Particularly, for an incident circular polarization, such a metasurface converts the incident light into a beam of opposite circular polarization, imprinted with a geometric phase $-2\sigma_{\pm}\theta(x,y)$, where $\sigma_{\pm} = \pm 1$ is the polarization helicity of the incident light corresponding to right and left circularly polarized light, respectively [6,9,19,20,25,26]. While the redirection of light by the metasurface is controlled by the imprinted geometric phase and the polarization state of the incident light (corresponding to the space-variant orientation angles), the metasurface efficiency is governed by the characteristics of the wave plate. Design-wise, to maximize the conversion efficiency of an incident circular polarization into the opposite circular polarization—cross-polarization efficiency (CrPE)—while minimizing the efficiency for maintaining the polarization—co-polarization efficiency (CoPE)—by the metasurface, the optical response of the meta-atom wave plate should mimic a half-wave plate, i.e., $t_x = t_y = 1$ and $\Phi = \pi$ [4,5,10]. As the meta-atoms, we considered ultrathin 1D poly-Si nanobeams with a thickness of 207 nm on a glass substrate, at the wavelength of 630 nm [see Fig. 5(c)]; the top-illuminated nanobeams are excited with polarizations perpendicular ($x$ direction) and parallel ($y$ direction) to the nanobeams grooves. By sweeping the width of the nanobeams and the period of the nanobeam array, we simulated the maps of $t_x$, $t_y$, and $\Phi$ [Figs. 5(d)–5(f)]. Figures 5(g) and 5(h) show the maps of the CoPE and CrPE $\left|\frac{1}{2}\left(t_x \pm t_y e^{i\Phi}\right)\right|^2$, respectively. We chose the width and period of the nanobeams according to the criterion of minimum CoPE, resulting in an optimized nanobeam width of 75 nm and a period of 230 nm [see the marked point in Figs. 5(g) and 5(h)], corresponding to an extremely low CoPE of 0.0015% and a considerably high CrPE of 77%. Note that the maximum CrPE and the minimum CoPE cannot be achieved simultaneously. To account for the highest possible CrPE while minimizing the CoPE, we utilized the degree of freedom of the nanobeam thickness. Figure 5(i) shows the dependence of the minimum CoPE on the nanobeam thickness; note that each data point was extracted from a CoPE map, in which the nanobeam width and period were swept—resulting in different combinations of beam width and period for each thickness. Moreover, for these thickness-dependent combinations of beam width and period, we introduced the dependence of the CrPE, and its difference with respect to the minimum CoPE, on the nanobeam thickness [Fig. 5(i)]. This analysis clearly shows that the nanobeam thickness of 207 nm yields the highest CrPE and the highest difference with respect to the minimum CoPE. The resulting structure parameters of this optimization process correspond to $t_x$, $t_y$, and $\Phi$ of 0.8733, 0.8808, and $0.9992\pi$, demonstrating a near-ideal half-wave plate design by the nanobeams. The reported multi-parameter optimization of the nanobeam-based wave plate paves the way for the optimized design of a metasurface based on the geometric phase.

### 3.3. Dammann metasurfaces based on the geometric phase

We realized DMs based on the Pancharatnam–Berry geometric phase by tiling the poly-Si nanobeams according to the target phase maps [Figs. 3(a)–3(c), 3(g) and 3(h)]; particularly, as the target is a binary phase $(0,\pi)$, the resulting orientation angle profile $\theta(x,y)$ is binary with angles 0 or $\pi/2$ [see Fig. 6(a) inset]. Figures 6(a)–6(c) depict the 1D DMs (unit cells) corresponding to 5, 7, and 9 equal-power diffraction orders, respectively, whereas Figs. 6(d)–6(f) present the corresponding simulated far-field patterns for an incident right circular polarization (RCP; $\sigma_+ = +1$) (all other simulation parameters are the same as in DG simulations). Note that the far-field patterns correspond to the cross-polarization component—opposite circular polarization with respect to the polarization excitation—i.e., left circular polarization (LCP; $\sigma_- = -1$). Similarly, Figs. 6(g) and 6(h) show the 2D DMs (unit cells) corresponding to 7-by-7 and 9-by-9 desired diffraction orders, respectively, where Figs. 6(i) and 6(j) exhibit the corresponding simulated far-field patterns [see Fig. 6(k) for the cross sections]. Uniformity-wise, DMs outperform the capabilities of DGs—see Figs. 7(a) and 7(b) for the thorough comparison. As for 1D structure realizations, DGs [Figs. 4(d)–4(f)] and DMs [Figs. 6(d)–6(f)] show high uniformities exceeding 90% and 96%, respectively. In stark contrast, 2D DMs remarkably outperform their DG counterparts by exhibiting high uniformities of 96%, 90%, and 84%, while 2D DGs show significantly lower uniformities of 66%, 55%, and 25%, for 5-by-5, 7-by-7, and 9-by-9 desired diffraction orders, respectively [Fig. 7(b)]. Moreover, the calculated RMSEs reveal a similar observation by exhibiting lower values for DMs in comparison with DGs [Figs. 2(i) and 2(j)], thus indicating the robust and highly precise phase imprint of DMs as the structure realization for the target phase. Notably, DMs overcome the uniformity defects in 2D DG diffraction patterns by solving the defective spatial phase imprint, thus outperforming the capabilities of DGs as beam multipliers.

Metasurfaces based on the Pancharatnam–Berry geometric phase are fundamentally manifested by a polarization-dependent response; however, here, the binary phase $(0,\pi)$ profile and the mirror symmetry between the desired diffraction orders yield DMs with indistinguishable phase profiles for an incident RCP or LCP, thus exhibiting a polarization-independent response [4]. While the simulations of DMs (Fig. 6) correspond to incident RCP, Fig. 7(c) shows the simulated far-field pattern for the orthogonal incident polarization, i.e., LCP. The far-field patterns corresponding to incident right [Fig. 2(h)] and left circularly polarized light [Fig. 7(c)] are identical in terms of the uniformity and diffraction efficiency, thus revealing the polarization-independent response of binary phase $(0,\pi)$ DMs. Figure 7(d) shows the far-field pattern corresponding to the co-polarization component which exhibits an extremely low intensity—resulting from the minimized CoPE by the metasurface design. This result implies that the far fields can be experimentally observed without polarization-resolved measurements [1,4,6,10,19], which significantly simplifies the experiments. To study the spectral behavior of DMs, we characterized the uniformity and diffraction efficiency over a wide wavelength range [Fig. 7(e)]. While for DMs, the dependence of the uniformity on the wavelength exhibits a quite uniform behavior with uniformities exceeding 88% over a bandwidth of 140 nm, DGs show a nonuniform strongly oscillating behavior [Fig. 7(e)]. Moreover, within this wavelength range, DMs exhibit a uniform diffraction efficiency of ~60%, which show good agreement with theoretical calculations of 2D DGs [35,37,42]. Strikingly, these observations reveal that DMs outperform their DG counterparts by exhibiting a broadband polarization-independent behavior.

### 3.4. The role of phase imprint imperfections on the uniformity defects in the diffraction patterns

The uniformity defects in the far-field diffraction patterns of 2D DGs emerge from the defective spatial phase imprint differing from the target binary phase $(0,\pi)$ profile [Fig. 2(i)]—induced by the structure realization. To study and quantify the correspondence between the uniformity defects in the far fields and the defective spatial phase imprint, we derived an analytical modular model in which we quantified the uniformity among the diffraction orders based on a specific error source of the phase imprint (see Supplement 1). Remarkably, the reported model reveals the mechanism for the uniformity defects in 2D DG diffraction patterns by showing the role of the different defect types of the spatial phase imprint in the uniformity among the diffraction orders (Fig. S1).

## 4. Conclusions

While DMs have already been studied [40,51–57], their comparison to DGs in the context of structure realizations to a target phase profile has not been reported so far. Remarkably, this study closes this gap. We showed that 2D DG structures exhibit uniformity defects in their far-field diffraction patterns; by studying the imprinted phase, we identified the mechanism of the uniformity defects as a defective spatial phase imprint, differing from the target phase near the transition points with subwavelength smallest features. Accordingly, we tailored a physics-driven solution using metasurfaces, where we showed that DMs outperform the capabilities of DGs by overcoming the uniformity defects via a robust and highly precise phase imprint. Moreover, this study has a contrast compared to previous studies in the field of DMs [40,51–57], using meta-atoms of anisotropic nanorods (nanofins) or isotropic circular or square nanopillars, while we use anisotropic nanogratings (periodic nanobeams). The latter enables full area coverage of the different regions in the phase maps [see Fig. 6(a) inset], giving rise to a precise determination of the transition points, which is crucial for obtaining equal-power diffraction orders. Therefore, nanograting-based metasurfaces [4,5,10,14] hold the promise for DMs with high uniformities.

Through their high uniformity, the reported DMs with equal-power diffraction orders provide distinct advantages in three-dimensional imaging based on structured light [58] via metadevices [59]. An array of dots with equal optical power provides a uniform signal-to-noise ratio across the field of view, thereby ensuring consistent spot localization precision and spatially homogenous depth accuracy (in three-dimensional imaging based on triangulation, the localization precision varies with the intensity). By avoiding intensity variations that can lead to saturation of bright spots and noise-limited detection of weak ones, equal-power projection maximizes the effective use of the detector dynamic range and improves the robustness of spot detection algorithms (e.g., by reducing false matches and missed detections and enabling simpler segmentation). This uniformity simplifies calibration and system modeling, as variations in measured response can be attributed primarily to scene reflectivity and geometry rather than illumination nonuniformity. Therefore, DMs exhibiting a remarkably high uniformity open a new paradigm for three-dimensional imaging applications based on structured light—such as facial recognition and light detection and ranging—via metadevices.

In summary, we report DMs based on the geometric phase which outperform the performance of conventional binary phase DGs by overcoming the uniformity defects in 2D DG diffraction patterns. We revealed that the low uniformity among the diffraction orders of 2D DG structures results from a defective spatial phase imprint that differs from the target phase near the transition points with subwavelength smallest features. We showed that 2D DMs exhibit high uniformity and diffraction efficiency, in contrast to DGs, by overcoming the uniformity defects via a robust and highly precise phase imprint. Strikingly, DMs with equal-power diffraction orders outperform their DG counterparts by exhibiting a polarization-independent response and a broadband operation. Moreover, while DGs are manifested by a wavelength-scale thickness but subwavelength smallest features, fabrication-wise, ultrathin DMs facilitate the realization by overcoming the challenge of high aspect ratio etching. While DMs empower binary phase DGs, the versatile metasurface platform holds the promise for other beam-multiplier structures. In comparison with DGs with a binary phase profile, DGs with a continuous phase reveal higher efficiencies while maintaining the high uniformity among the diffraction orders [35,37,60]. As such, DMs provide a route to realize continuous phase DGs in an optically thin flatland, overcoming the arduous task of fabricating optical elements with a continuously varying thickness. Broadband DMs with near-unity uniformity and efficiency open a new paradigm for three-dimensional imaging applications based on structured light, such as facial recognition, and beam combiners for high-power laser devices. Notably, the reported DMs can be fabricated using standard electron beam lithography followed by lift-off processes. Experiment-wise, far-field measurements can be obtained by illuminating the metasurface with a laser at the operating wavelength and imaging the $k$ space of the transmitted light—by imaging the back focal plane of the objective lens capturing the metasurface-scattered light onto a camera. Remarkably, this study redefines the optical design paradigms for binary phase optical elements revealing that metasurfaces can outperform the capabilities of their bulk optics counterparts while facilitating virtually flat, ultrathin, and lightweight optics.

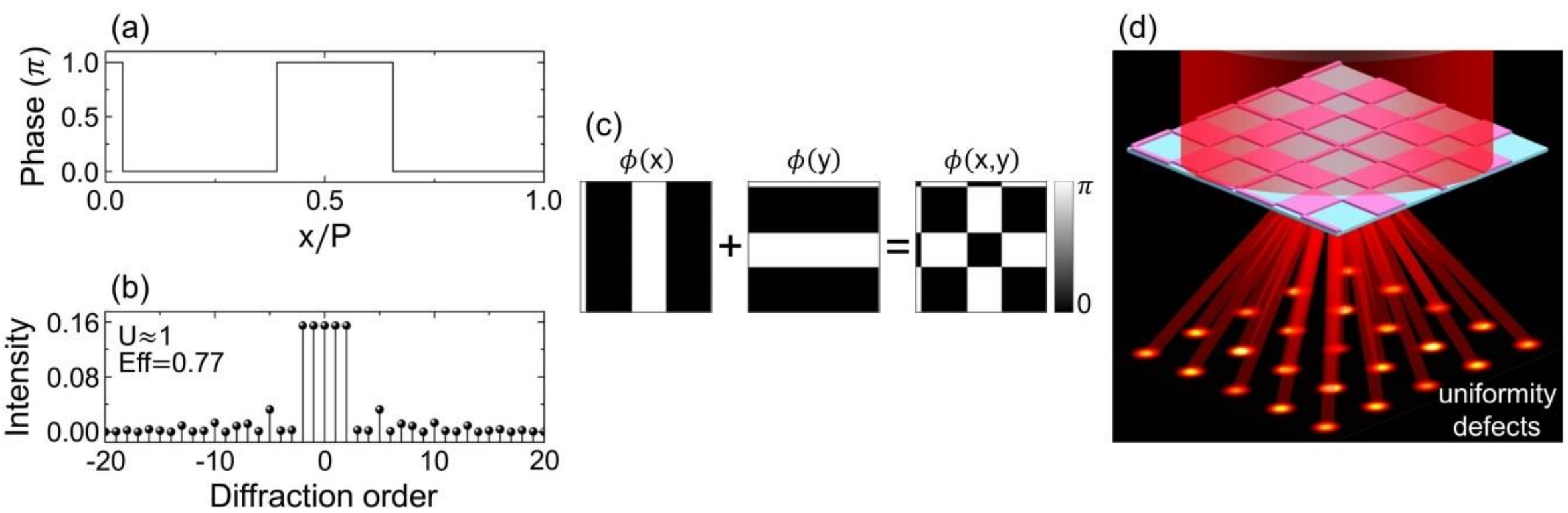

**Fig. 1.** Dammann grating operation as a beam multiplier: Working principles and uniformity defects in two-dimensional diffraction patterns. (a) Typical binary phase $(0,\pi)$ profile of a 1D DG with the carefully designed transition points within the period of the grating $P$; specifically, this design generates 5 equal-power diffraction orders. (b) Intensities ($|c_n|^2$) of the angular spectrum revealing the near-unity uniformity of the desired diffraction orders along with their high efficiency (Eff). (c) Procedure for obtaining 2D DGs with barcode-like patterns by summing two 1D DG phase maps in the horizontal and vertical directions. (d) Schematics of the DG operation as a beam multiplier, where the 2D diffraction pattern exhibits uniformity defects.

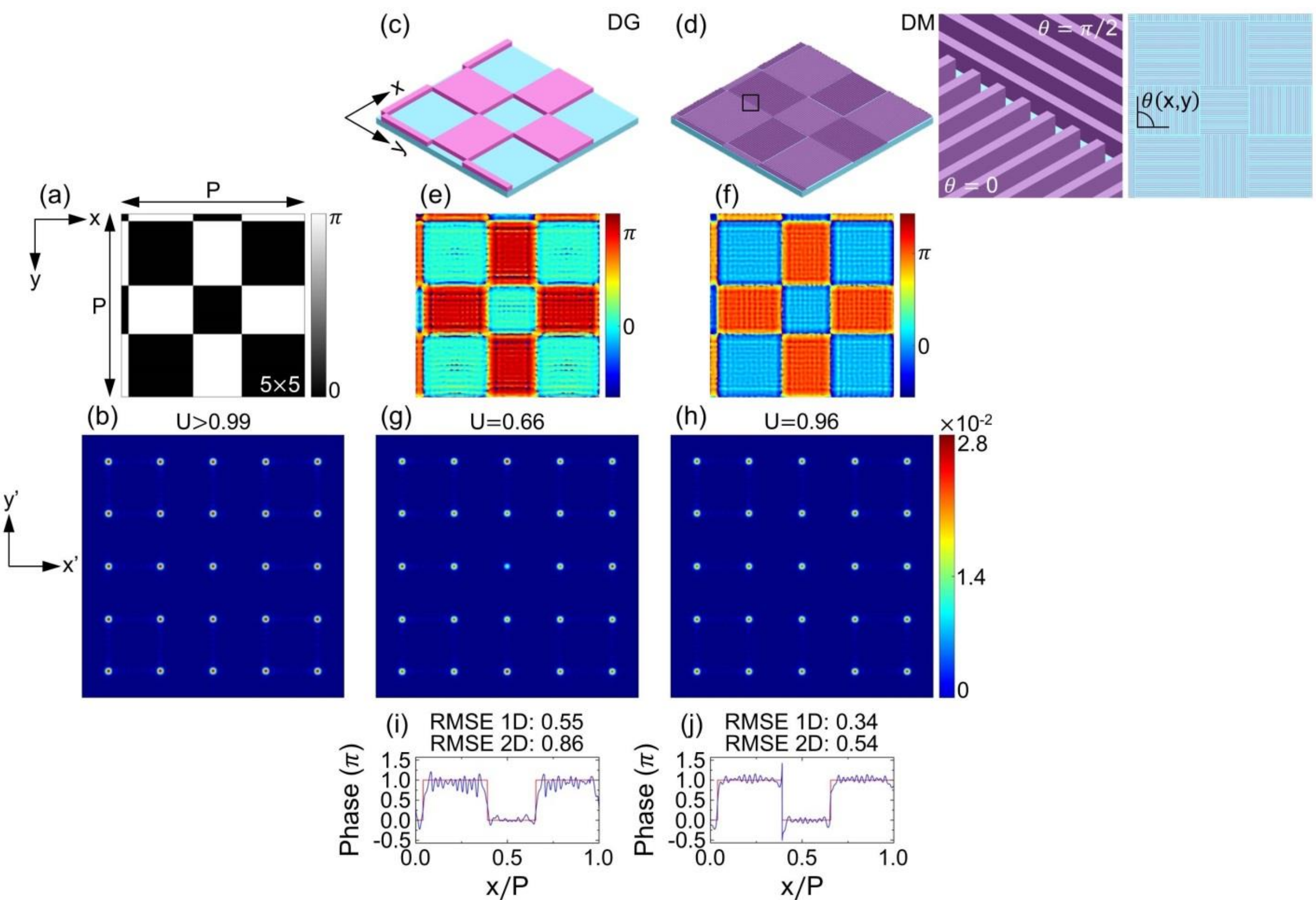

**Fig. 2.** Dammann metasurface route to overcoming the uniformity defects in two-dimensional beam multipliers. (a) Typical binary phase $(0,\pi)$ map (single period) of a 2D DG; specifically, this design generates 5-by-5 equal-power diffraction orders. (b) Calculated far-field intensity pattern based on the perfect phase profile. Here, the desired diffraction orders are extremely uniform, with a calculated uniformity exceeding 99%—where no structure realization for the target phase is considered. (c) Typical structure realization (single period) of a 2D binary phase $(0,\pi)$ DG structure; based on the target binary phase $(0,\pi)$ profile, this traditional structure is obtained by selectively etching the grating material to a thickness corresponding to a $\pi$ phase shift. (d) Typical structure realization (single period) of a 2D binary phase $(0,\pi)$ DM; based on the target phase profile, the DM is obtained by tiling anisotropic nanogratings according to an on-demand space-variant orientation angle profile $\theta(x,y)$—see the insets of a zoom-in image and the top view of the DM. (e), (f) Simulated near-field phase profiles of the DG structure and DM, respectively. (g), (h) Simulated far-field intensity patterns of the DG structure and DM, respectively. (i), (j) Typical Horizontal cross sections of the DG and DM imprinted and target phases. The 1D and 2D RMSE calculations correspond to the cross section and map data,

respectively. The structure realization for the target phase of a 2D DG is manifested by a low uniformity (66%) among the diffraction orders in its far-field pattern, induced by a defective spatial phase imprint with high RMSEs. Strikingly, the structure realization of a DM overcomes the uniformity defects in the 2D diffraction pattern—with a remarkably high uniformity of 96%—by a highly precise phase imprint that features low RMSEs.

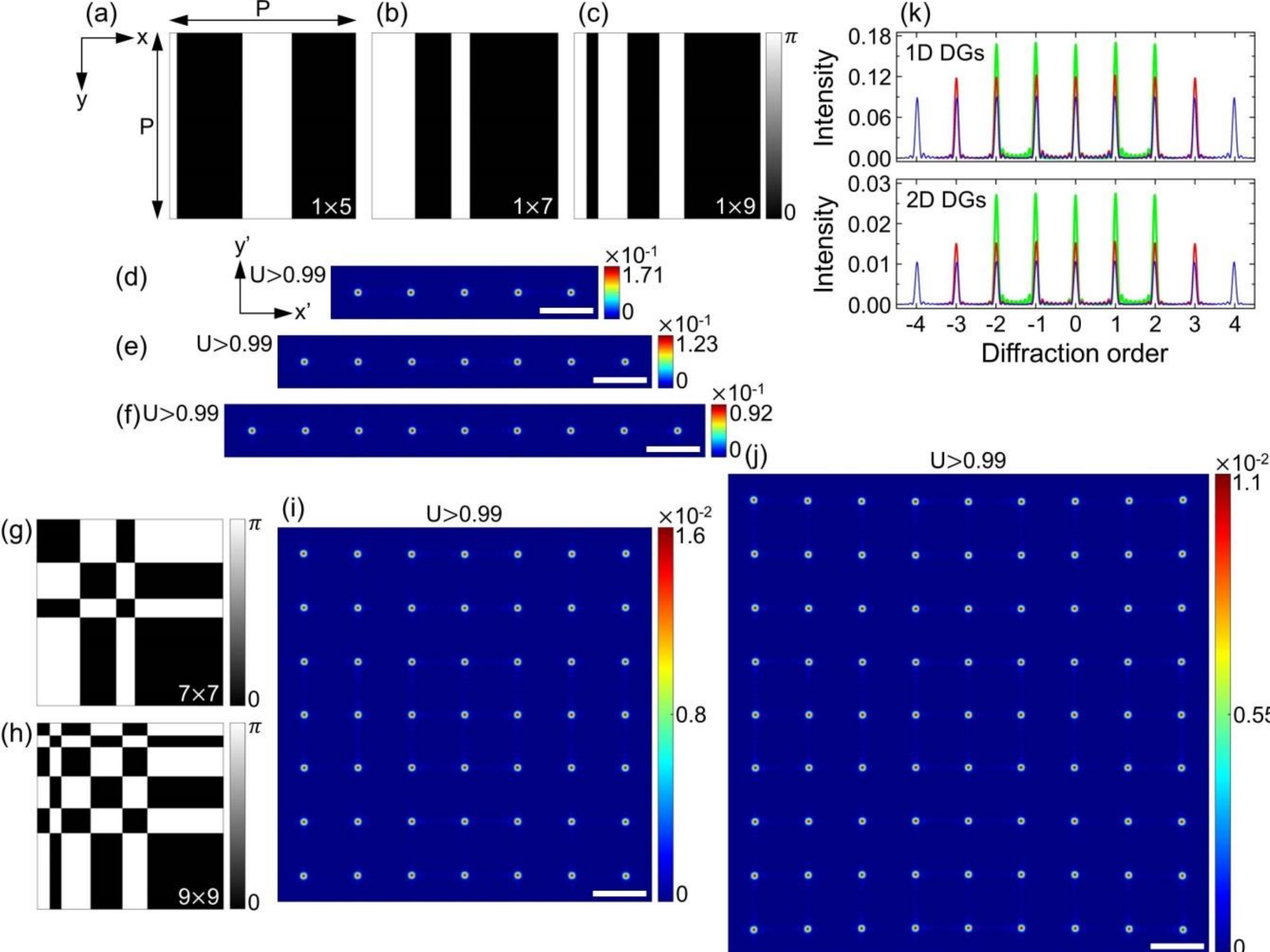

**Fig. 3.** Far-field patterns of Dammann gratings based on the perfect phase profile. (a)–(c) 1D binary phase $(0,\pi)$ maps corresponding to 5, 7, and 9 equal-power diffraction orders, respectively; the maps correspond to a single unit cell of the periodic DGs with a period $P$ of 20 μm. (d)–(f) Corresponding calculated far-field intensity patterns; the calculation considers DGs with a total size of 200 μm, at the wavelength of 630 nm. The scale bars correspond to a distance between two adjacent diffraction orders at a far-field distance $z$ of 22 m. (g), (h) 2D binary phase $(0,\pi)$ maps (unit cells) corresponding to 7-by-7 and 9-by-9 desired diffraction orders, respectively. (i), (j) Corresponding calculated far-field intensity patterns. (k) Horizontal cross sections of the intensity distributions at the center of the far-field maps corresponding to 5, 7, and 9 (1D DGs) and 5-by-5, 7-by-7, and 9-by-9 (2D DGs) equal-power diffraction orders; in 2D DGs, the vertical and horizontal cross sections are identical. These theoretical calculations based on the perfect phase profile reveal the extremely uniform diffraction orders of 1D and 2D DGs, with uniformities exceeding 99%. Note that this calculation disregards a structure realization for the target phase.

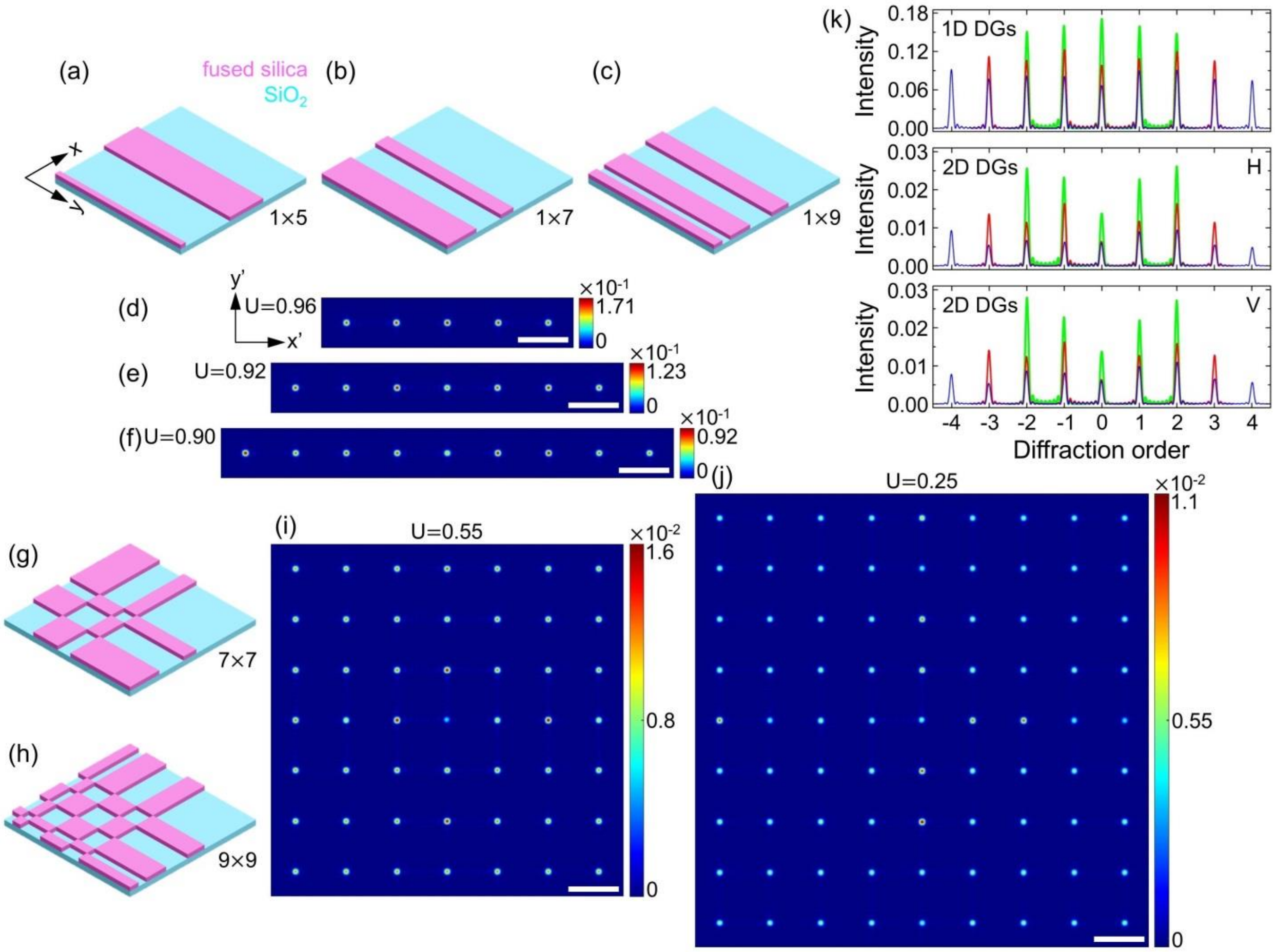

**Fig. 4.** Dammann grating structures based on a structure realization of selective etching: Far-field patterns. (a)–(c) 1D binary phase $(0,\pi)$ DG structures (unit cells) corresponding to 5, 7, and 9 equal-power diffraction orders, respectively. According to the target binary phase $(0,\pi)$ profiles, these traditional structures are obtained by selectively etching the grating material—fused silica—to a thickness corresponding to a $\pi$ phase shift, at the wavelength of 630 nm (a glass ($SiO_2$) substrate is considered). Similar to the DG calculations, the period and total size of the simulated DG structures are 20 μm and 200 μm, respectively. (d)–(f) Corresponding simulated far-field intensity patterns; the simulation considers a normally incident horizontal linear polarization, where the polarization state of the far field is the same as the polarization excitation. The scale bars correspond to a distance between two adjacent diffraction orders at a far-field distance of 1 m. (g), (h) 2D binary phase $(0,\pi)$ DG structures (unit cells) corresponding to 7-by-7 and 9-by-9 desired diffraction orders, respectively. (i), (j) Corresponding simulated far-field intensity patterns. (k) Horizontal (H) and vertical (V) cross sections of the intensity distributions at the center of the far-field maps corresponding to 5, 7, and 9 (1D DGs) and 5-by-5, 7-by-7, and 9-by-9 (2D DGs) equal-power diffraction orders. While the far-field patterns of 1D DG structures exhibit high uniformities (exceeding 90%), the 2D DG structures show extremely low uniformities (as low as 25%), revealing that the structure realization for the target phase plays a crucial role in DG performances.

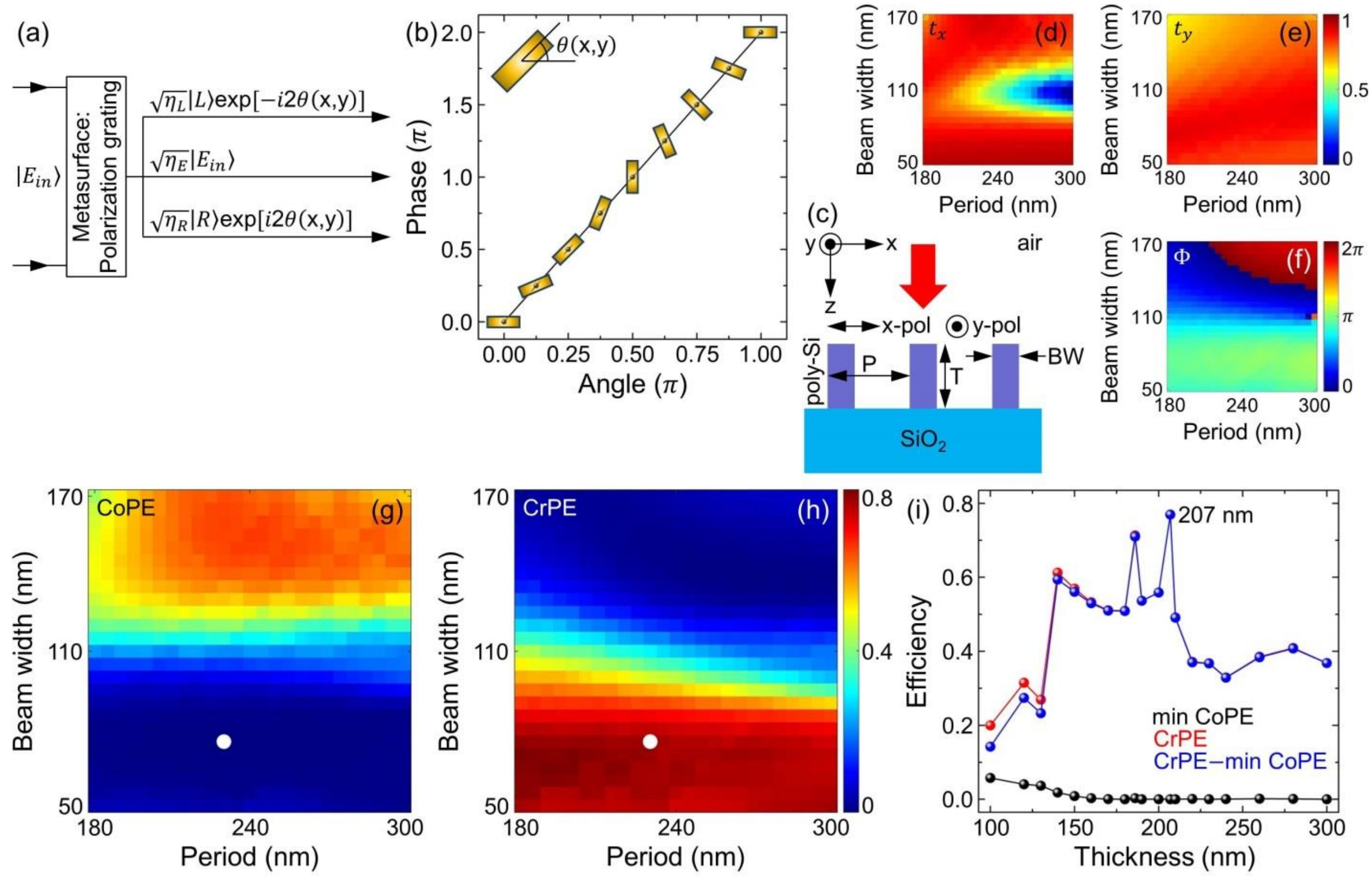

**Fig. 5.** Metasurface based on the Pancharatnam–Berry geometric phase: Operation principles and design method. (a) Diagram describing the operation of a linear metasurface based on the Pancharatnam–Berry geometric phase with a space-variant orientation angle profile $\theta(x,y)$. An incident beam with a polarization $|E_{in}\rangle$ excites the metasurface. The resulting beam comprises three polarization orders: $|E_{in}\rangle$ polarization order which maintains the polarization and phase of the incident beam, and $|R\rangle$ and $|L\rangle$ polarization orders which are accompanied by geometric phase modifications of $\pm 2\theta(x,y)$, respectively. (b) Geometric phase pickup of $2\theta$, where $\theta$ is the orientation angle of the anisotropic meta-atom. (c) Schematic view of a periodic poly-Si nanobeam array on a glass ($SiO_2$) substrate. The top-illuminated nanobeams are excited with polarizations perpendicular ($x$ direction) and parallel ($y$ direction) to the nanobeams grooves. (d)–(f) Maps of $t_x$, $t_y$, and $\Phi$ as a function of the beam width (BW) and period of the nanobeam array, at the wavelength of 630 nm. (g), (h) CoPE and CrPE maps as a function of the nanobeam width and period. The marked point corresponds to the minimum CoPE with an optimized nanobeam width of 75 nm and a period of 230 nm. (i) Dependence of the minimum CoPE, CrPE, and their difference on the nanobeam thickness. Here, for each thickness, the beam width and period are different as they were extracted from a CoPE map (in which the nanobeam width and period were swept) as the minimum value. This dependence shows that the optimized nanobeam thickness is 207 nm.

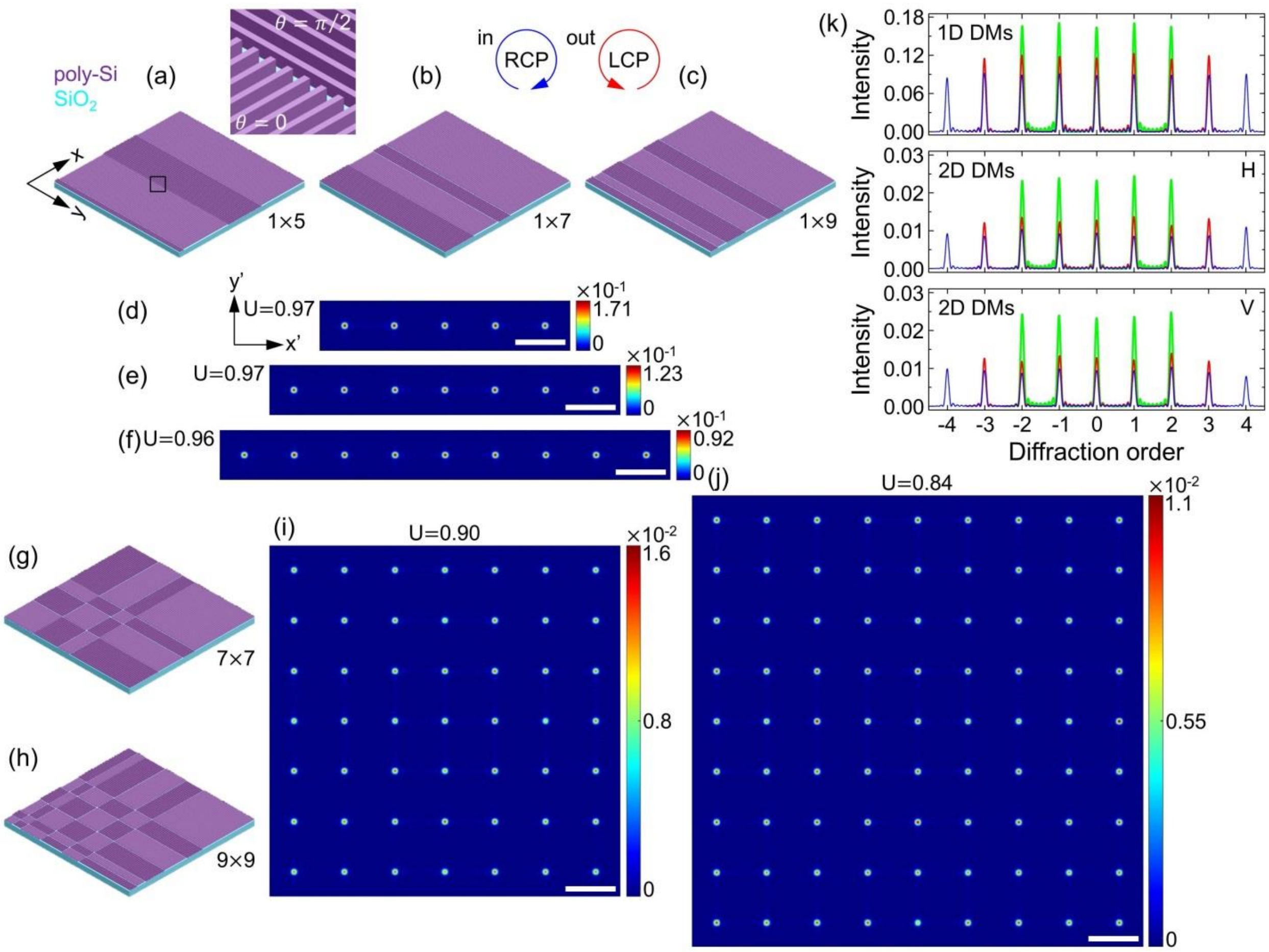

**Fig. 6.** Dammann metasurfaces based on a structure realization of the geometric phase: Far-field patterns. (a)–(c) 1D binary phase $(0,\pi)$ DMs (unit cells) corresponding to 5, 7, and 9 equal-power diffraction orders, respectively. Based on the target binary phase $(0,\pi)$ profiles $\phi(x,y)$, the DMs are obtained by tiling the poly-Si nanobeams according to a binary orientation angle profile $\theta(x,y) = \phi(x,y)/2$—with angles 0 or $\pi/2$ (see the inset of a zoom-in image). The width, period, and thickness of the poly-Si nanobeams are 75 nm, 230 nm, and 207 nm, respectively. The grating period and the total size of the simulated DMs are the same as in the DG simulations. (d)–(f) Corresponding simulated far-field intensity patterns; the simulation considers a normally incident RCP, at the wavelength of 630 nm, where the polarization of the far field is LCP (opposite circular polarization with respect to the polarization excitation). (g), (h) 2D binary phase $(0,\pi)$ DMs (unit cells) corresponding to 7-by-7 and 9-by-9 desired diffraction orders, respectively. (i), (j) Corresponding simulated far-field intensity patterns. (k) Horizontal and vertical cross sections of the intensity distributions at the center of the far-field maps corresponding to 5, 7, and 9 (1D DGs) and 5-by-5, 7-by-7, and 9-by-9 (2D DGs) equal-power

diffraction orders. Uniformity-wise, 1D and 2D DMs outperform the capabilities of DGs; particularly, 2D DMs remarkably outperform their DG counterparts by exhibiting significantly higher uniformities.

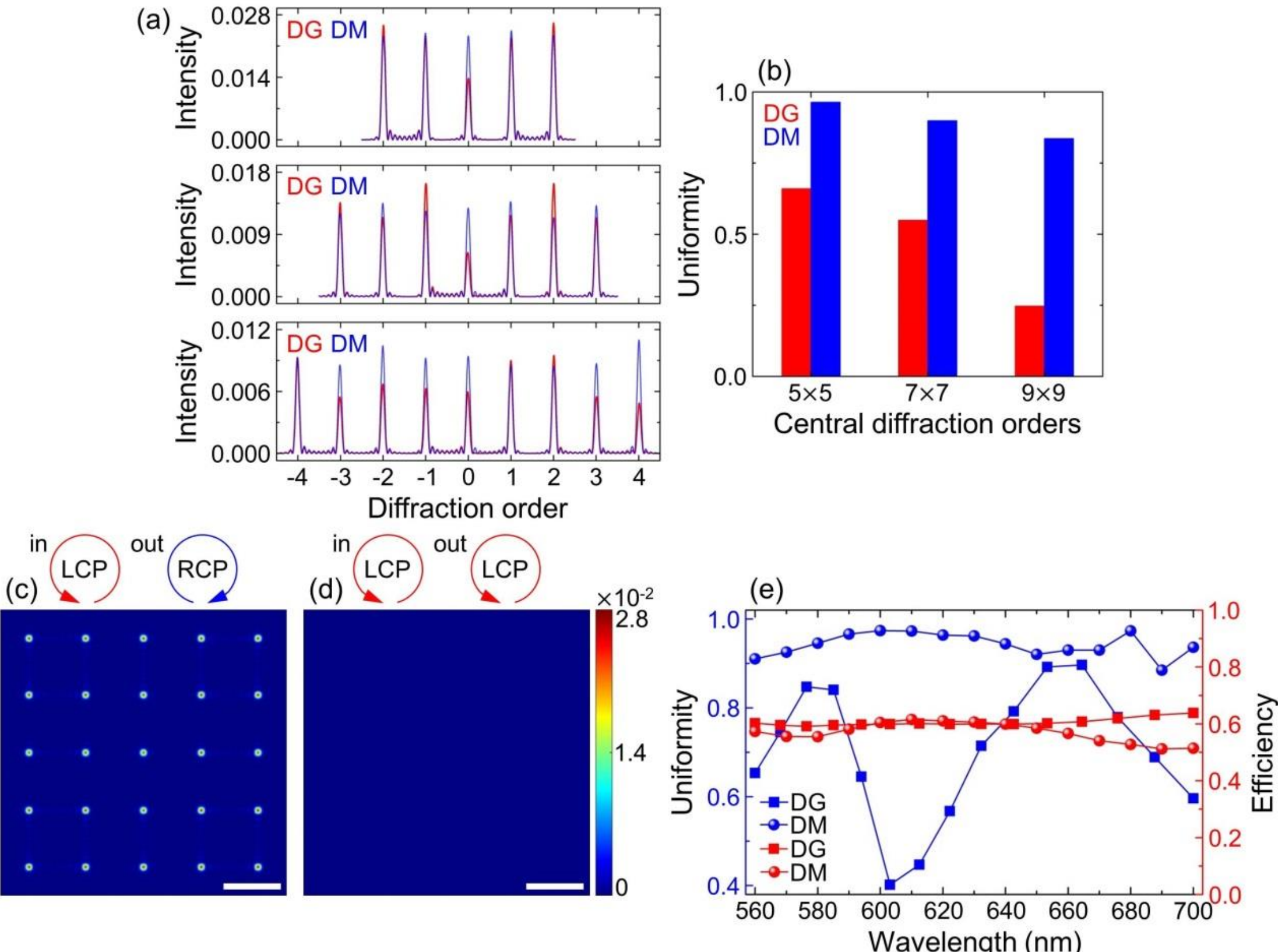

**Fig. 7.** Polarization-independent and broadband response of Dammann metasurfaces. (a) Comparison of the (horizontal) cross sections of the far-field intensities between 2D DGs and DMs corresponding to 5-by-5, 7-by-7, and 9-by-9 equal-power diffraction orders; in contrast to DGs, the DM cross sections exhibit high uniformities. (b) Uniformity comparison of 2D DGs and DMs which shows the significantly higher uniformities of DMs. (c) Far-field intensity map of the cross-polarization component of a 2D DM corresponding to 5-by-5 desired diffraction orders for an incident LCP; this result reveals the polarization-independent response of binary phase $(0,\pi)$ DMs based on the geometric phase. (d) Corresponding far-field intensity map of the co-polarization component, revealing the minimized CoPE by the metasurface design. (e) Dependence of the uniformity and diffraction efficiency on the wavelength for DGs and DMs. DMs outperform their DG counterparts by exhibiting a uniform behavior of the uniformity over a wide wavelength range.

Supplement 1 includes an analytical model: from a defective spatial phase imprint to uniformity defects in the far-field diffraction pattern.

**Acknowledgments**

The authors gratefully acknowledge funding from the Israel Science Foundation (ISF) under Grant No. 1785/22.

**Disclosures**

The authors declare no conflicts of interest.

**Data availability**

Data underlying the results presented in this paper are not publicly available at this time but may be obtained from the authors upon reasonable request.

Supplemental Document for

# Dammann metasurface route to overcoming the uniformity defects in two-dimensional beam multipliers

Rinat Gutin[1†], Raghvendra P. Chaudhary[1†], Imon Kalyan[1], Avraham Reiner[1], and Nir Shitrit[1*]

[1]*School of Electrical and Computer Engineering, Ben-Gurion University of the Negev, Be'er Sheva 8410501, Israel*

[†]These authors contributed equally to this work.

[*]Corresponding author. E-mail: nshitrit@bgu.ac.il

**From a defective spatial phase imprint to uniformity defects in the far-field diffraction pattern: An analytical model**

The uniformity defects in the far-field diffraction patterns of two-dimensional (2D) Dammann gratings (DGs) emerge from the defective spatial phase imprint differing from the target binary phase $(0,\pi)$ profile [Fig. 2(i)]—induced by the structure realization. The observed imperfections are identified as phase jumps near the transition points with a maximum phase value exceeding $\pi$ and a finite decay length, and a quasi-binary phase (0,non-$\pi$) profile [see Fig. 2(i)]. To study and quantify the correspondence between the uniformity defects in the far fields and the defective spatial phase imprint, we derived an analytical modular model in which we calculated the far-field patterns and quantified the uniformity among the diffraction orders based on a specific error source of the phase imprint. Particularly, we consider the cases of (i) a binary phase $(0,\pi+\varepsilon)$ profile, where $\varepsilon$ represents the difference from the perfect binary phase $(0,\pi)$ profile [Fig. S1(a) inset], (ii) Lorentzian-shaped phase jumps in the transition points with varying Lorentzian widths $\gamma$ [Fig. S1(b) inset], and (iii) varying peak values $\phi_{\max}$ [Fig. S1(c) inset]. Moreover, to characterize the uniformity defects, we calculated the uniformity including all diffraction orders and excluding the zero order; this approach allows to study the role of the phase imprint imperfections in the uniformity of either all diffraction orders or the zero order. We revealed that the defect of a binary phase (0,non-$\pi$) profile gives rise to a reduction of the uniformity primarily owing to modifications in the zero-order intensity [Fig. S1(a)]; this observation originates from similar uniformity trends corresponding to the one-dimensional (1D) DG and 2D DG excluding the zero order (referred to as 2D*), while the uniformity trend corresponding to a 2D DG shows significantly lower uniformities. In stark contrast, the dependences of the uniformity on the width [Fig. S1(b)] and the peak value of the phase jumps in the transition points [Fig. S1(c)] give rise to a uniformity reduction across all diffraction orders. While the uniformity defects in the far-field diffraction patterns of 2D DGs were observed in many experiments [38,39,43–46], their mechanism has remained elusive; remarkably, the reported model reveals the mechanism for the uniformity defects in 2D DG diffraction patterns by showing the role of the different defect types of the spatial phase imprint in the uniformity among the diffraction orders.

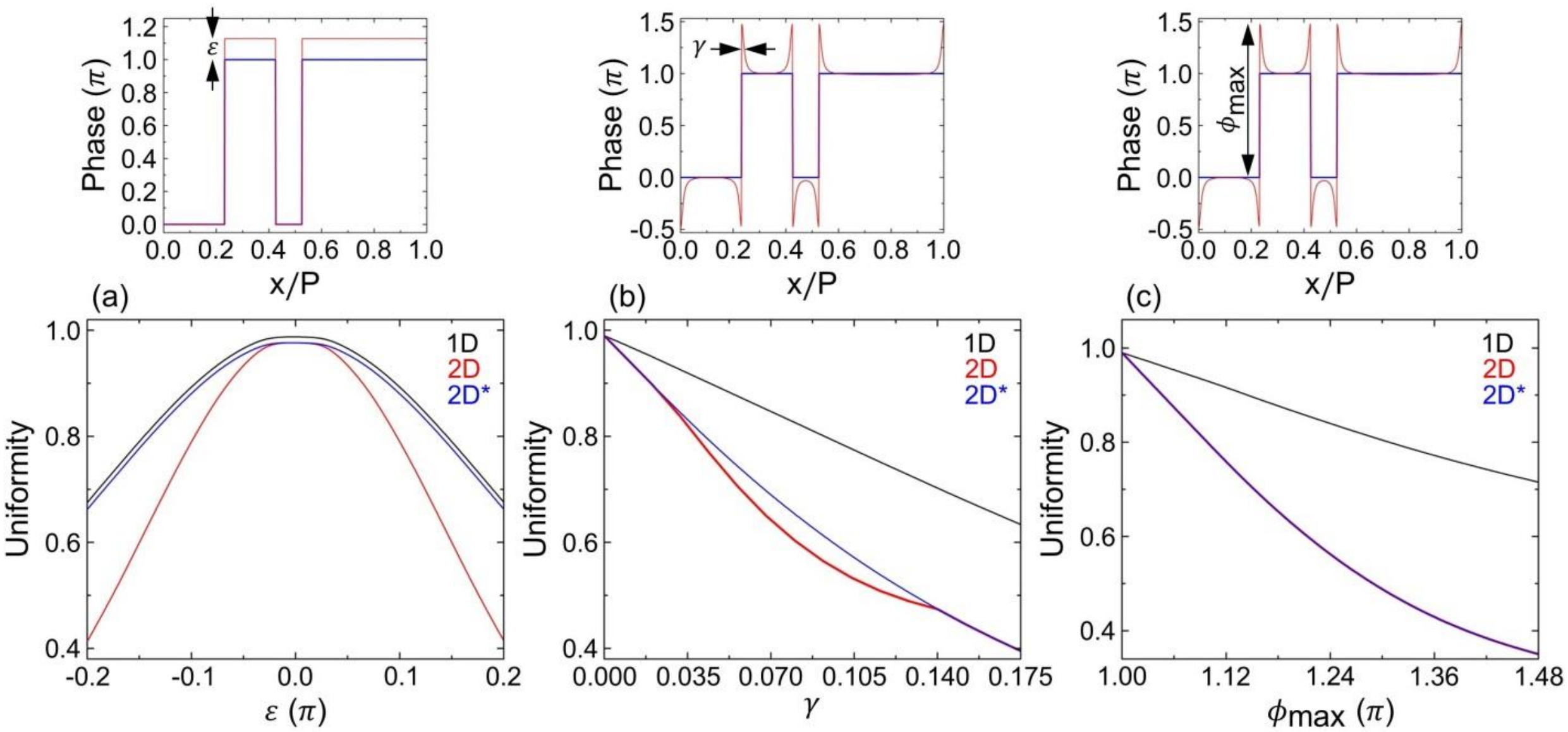

**Fig. S1.** Effects of different phase imprint imperfections on the uniformity of the far-field diffraction pattern: A theoretical model. (a) The case of a binary phase (0,non-$\pi$) profile. The inset shows the binary phase $(0,\pi+\varepsilon)$ profile of a 1D DG corresponding to 7 equal-power diffraction orders, with the carefully designed transition points within the period of the grating $P$; $\varepsilon$ represents the difference from the perfect binary phase $(0,\pi)$ profile. The figure shows the dependence of the uniformity on the phase imperfection $\varepsilon$. The calculated uniformities for the 1D DG and 2D DG excluding the zero order (2D*) are similar, while the uniformity corresponding to the 2D DG exhibit significantly lower values; this observation reveals that the phase imperfection $\varepsilon$ gives rise to a reduction of the uniformity primarily owing to modifications in the zero-order intensity. (b) The case of Lorentzian-shaped phase jumps in the transition points with varying Lorentzian widths $\gamma$. The inset shows the perturbed phase, where $\gamma$ represents the half width at half maximum of the Lorentzian. The figure exhibits the dependence of the uniformity on the phase imperfection $\gamma$. (c) The case of Lorentzian-shaped phase jumps in the transition points with varying peak values $\phi_{\max}$. The inset shows the perturbed phase, and the figure exhibits the dependence of the uniformity on the phase imperfection $\phi_{\max}$. In contrast to the case of a binary phase (0,non-$\pi$) profile, the cases of phase imperfections induced by phase jumps in the transition points with varying widths or peak values show similar uniformity trends for 2D and 2D* configurations—revealing that the uniformity reduction is across all diffraction orders.